\documentstyle[epsfig,12pt]{article}
\def\la{\hbox{\rlap{\raise.3ex\hbox{$<$}}\lower.8ex\hbox{$\sim$}\ }}
\def\ga{\hbox{\rlap{\raise.3ex\hbox{$>$}}\lower.8ex\hbox{$\sim$}\ }}

\begin{document}

\pagestyle{empty}

\centerline{\Large PROTOTYPE IMAGING CD-ZN-TE ARRAY DETECTOR}

\vspace{4mm}
\noindent P. F. BLOSER$^{1}$, T. NARITA$^{1}$,
J. E. GRINDLAY$^{1}$, K. SHAH$^{2}$

\noindent $^{1}$Harvard-Smithsonian Center for Astrophysics,
Cambridge, MA 02138

\noindent $^{2}$Radiation Monitoring Devices, Inc., Watertown, MA 02172

\vspace{8mm}

\noindent ABSTRACT

\vspace{4mm}

We describe initial results of our program to develop 
and test Cd-Zn-Te (CZT) detectors with a pixellated array 
readout. Our primary interest is in the development of 
relatively thick CZT detectors for use in astrophysical 
coded aperture telescopes with response extending over the 
energy range $\sim 10-600$ keV. The coded aperture imaging 
configuration requires only relatively large area pixels 
(1-3 mm), whereas the desired high energy response requires 
detector thicknesses of at least 3-5 mm. We have developed 
a prototype detector employing a 10 x 10 x 5 mm CZT substrate 
and 4 x 4 pixel (1.5 mm each) readout with gold metal contacts 
for the pixels and continuous gold contact for the bias on 
the opposite detector face. This MSM contact configuration 
was fabricated by RMD and tested at Harvard for uniformity, 
efficiency and spatial as well as spectral resolution. We 
have developed an ASIC readout (IDE-VA-1) and analysis 
system and report results, including $\sim 4$\% (FWHM) energy
resolution at 60 keV.  A prototype design for 
a full imaging detector array is discussed.

\vspace{4mm}

\noindent INTRODUCTION

\vspace{4mm}

The field of X-ray and gamma-ray astronomy has been limited since its
beginnings by the somewhat primitive state of high energy radiation
detector technology, especially imaging technology.  Particularly in
the hard X-ray energy range ($\sim 20-500$
keV), creating imaging instruments is difficult, since energies are too
high for multi-layer focusing optics (\la 80 keV) and too low for Compton
telescopes (\ga 500 keV).  The optimum method for imaging in
this range is the coded aperture technique, in which a
specially-designed mask casts a shadow on a position-sensitive
detector from which the position of the X-ray source can be deduced.
Though successfully employed in many experiments, this method has been
limited by the
poor spatial resolution of the scintillation detectors (NaI, CsI,
etc.) used to date.  A solid-state detector with small pixels 
would improve the spatial resolution immensely.
Scintillators also have poor energy resolution
compared to solid-state, semiconductor devices.  Thus hard X-ray
astronomy is in need of large area, room temperature,
semiconducting imaging detectors with good energy resolution.

Recently, Cadmium-Zinc-Telluride (CZT) has shown great promise in
meeting this need.  Alloying CdTe with Zn increases resistivity [1],
allowing higher bias voltage without increased leakage 
current.  CZT has high stopping power
and a wide bandgap, but can suffer from poor charge collection due
to deep traps and poor mobility-lifetime ($\mu\tau$) products for
holes.  However, it has been found [2] that employing a grid
of pixels small relative to the detector thickness creates an internal
electric field favorable for the collection of only one polarity of charge
carrier; if used to collect electrons, the easily-trapped holes are no
longer important (this is often called the {\em small pixel effect}).
As this is precisely the detector geometry required for hard X-ray  
telescope detectors, pixellated arrays of CZT have enjoyed intense
scrutiny in recent years [3-6].  Here we describe initial results of our
program to develop imaging CZT array detectors specialized for use in
a hard X-ray survey telescope.  

\vspace{4mm}

\noindent EXPERIMENT

\vspace{4mm}

Our detector development work has been motivated by a hard X-ray
survey telescope concept such as the previously proposed MIDEX mission
EXIST [7].  The purpose of such a mission is to conduct an all-sky
survey with high sensitivity and resolution 
(spatial and spectral) from $\sim 10-600$ keV; therefore, the
telescope must have a
large field of view with thick ($\sim 5$ mm) CZT detector elements.
Both these considerations lead us to a pixellated detector design.  
Since relatively defect-free CZT crystals are currently most readily available
in $< 10-12$ mm sizes, we are led naturally to a unit detector element
of $12 \times 12 \times 5$ mm with a $4 \times 4$ array of 2.5 mm
pixels spaced by 0.5 mm.  These could be grouped into a $2 \times 2$
array, which we call a basic detector
element (BDE), and read out by a single 64 channel
preamp-shaper-multiplexer
Application Specific Integrated Circuit
(ASIC) with self-triggering and low power consumption.  These BDEs can
then be tiled into a large array. 

A different approach is a strip detector, with perpendicular strips on
opposite sides to record x and y positions from electrons and holes
[4][5].
While strip detectors
require fewer channels (2N) for readout than do pixel 
(N$^2$) detectors, they
have the disadvantage of requiring collection of both electrons and
holes to record x and y positions; this limits their thickness and
thus their high energy 
response. Also, strips are generally much longer than pixels and thus 
have higher capacitance and noise. 
Finally, the large field of view combined with the large
thickness requires large detector elements to minimize charge
spreading effects.
For example, for the $40^{\circ}$ field of view of EXIST and a 5 mm
thick detector, the pixels must be $> 5 {\rm mm} \times
\tan(20^{\circ}) = 1.8$ mm across to ensure the charge spreads no
further than the nearest neighbor pixel.  
Such large pixels allow us to cover a big collecting area 
with relatively few pixels which can be readily coupled to  
multi-channel ASIC readouts.

Therefore, we are conducting extensive development and testing of thick
pixellated CZT detectors and multi-channel ASICs.  Our immediate goal
is to
construct a BDE and fly it on a scientific balloon in order to measure
background and uniformity of response under space flight conditions.
As a part of this effort RMD 
Corp. has fabricated a $10 \times 10 \times 5$ mm detector out of high
pressure Bridgman counter grade CZT obtained from eV Products, depositing a $4
\times 4$ array of 1.5 mm gold pixels spaced by 0.2 mm on one side and
a continuous gold contact on the other. A 1.5 mm wide gold ``guard ring'' 
surrounds (0.2 mm gap) the pixel array and was maintained 
at ground potential.

\begin{figure}[t] 
\hspace{1.6in}\epsfig{file=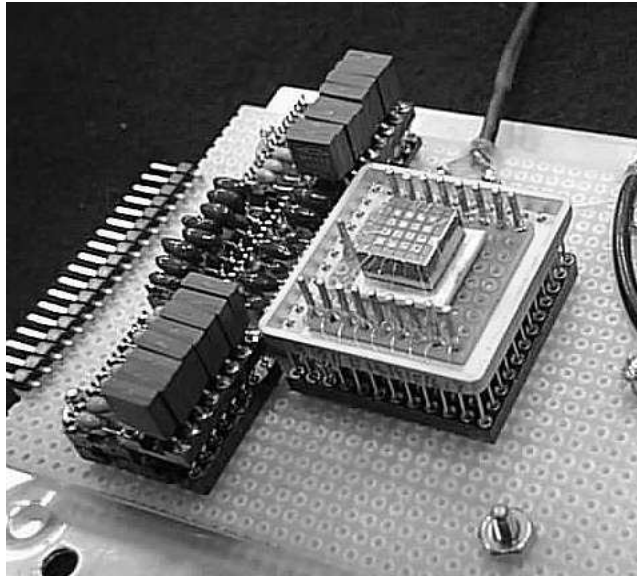,height=3in,width=3.3in}
\caption{The RMD CZT detector (10 $\times 10 \times$ 5 mm), 
with a $4 \times 4$ array
of 1.5 mm gold pixels, mounted in its metal-ceramic chip carrier 
with (to left) bias resistors and decoupling capacitors. The detector 
is illuminated from ``below'', through the chip carrier.}
\label{fig1}
\end{figure}

At CfA, we have tested this
detector for spatial and spectral resolution using a $^{241}$Am
source of 60 keV X-rays.  The detector was read out by two 8 channel
VA-1 preamplifier ASICs provided by IDE AS Corp. The CZT    
is irradiated through the metal-ceramic chip carrier on
the side with the continuous contact, 
which is negatively biased, so that the electrons drift nearly the entire 5
mm thickness to the grounded pixels.
Gold wires from the chip carrier holding the detector are connected to
the pixels with silver epoxy. 
The chip carrier
sits on a small board providing decoupling
capacitors and bias resistors for each channel (cf. Figure~\ref{fig1}); 
this board in turn
plugs into a test board provided by IDE AS Corp. that carries the
ASICs and provides line drivers for each output channel.  This board
was not optimized for minimal capacitance, as indicated below.  A lab pulser
was input through 1.4 pf capacitors into each ASIC channel in parallel
with the detector.  Spectra were then
recorded one channel at a time on a lab PC with a commercial MCA
card.  The MCA triggered internally on the shaped pulse; we also
experimented with deriving a trigger pulse from the cathode of the
detector (via a separate preamp) and got the same results.
The $^{241}$Am source was mounted on a computer-driven 
translation table allowing
collimated X-rays to be scanned across the detector with $\sim 0.2$ mm
resolution.  A photograph of
the RMD detector is shown in Figure~\ref{fig1}.

\vspace{4mm}

\noindent RESULTS

\vspace{4mm}

\begin{figure}[t] 
\begin{minipage}[t]{3.1in}
\epsfig{file=fig2.ps,height=2.1in,width=3in}
\caption{Linearity of VA-1 ASIC.  This input voltage range corresponds
to 30-180 keV.}
\label{fig2}
\end{minipage}
\hspace*{0.2in}
\begin{minipage}[t]{3.1in}
\epsfig{file=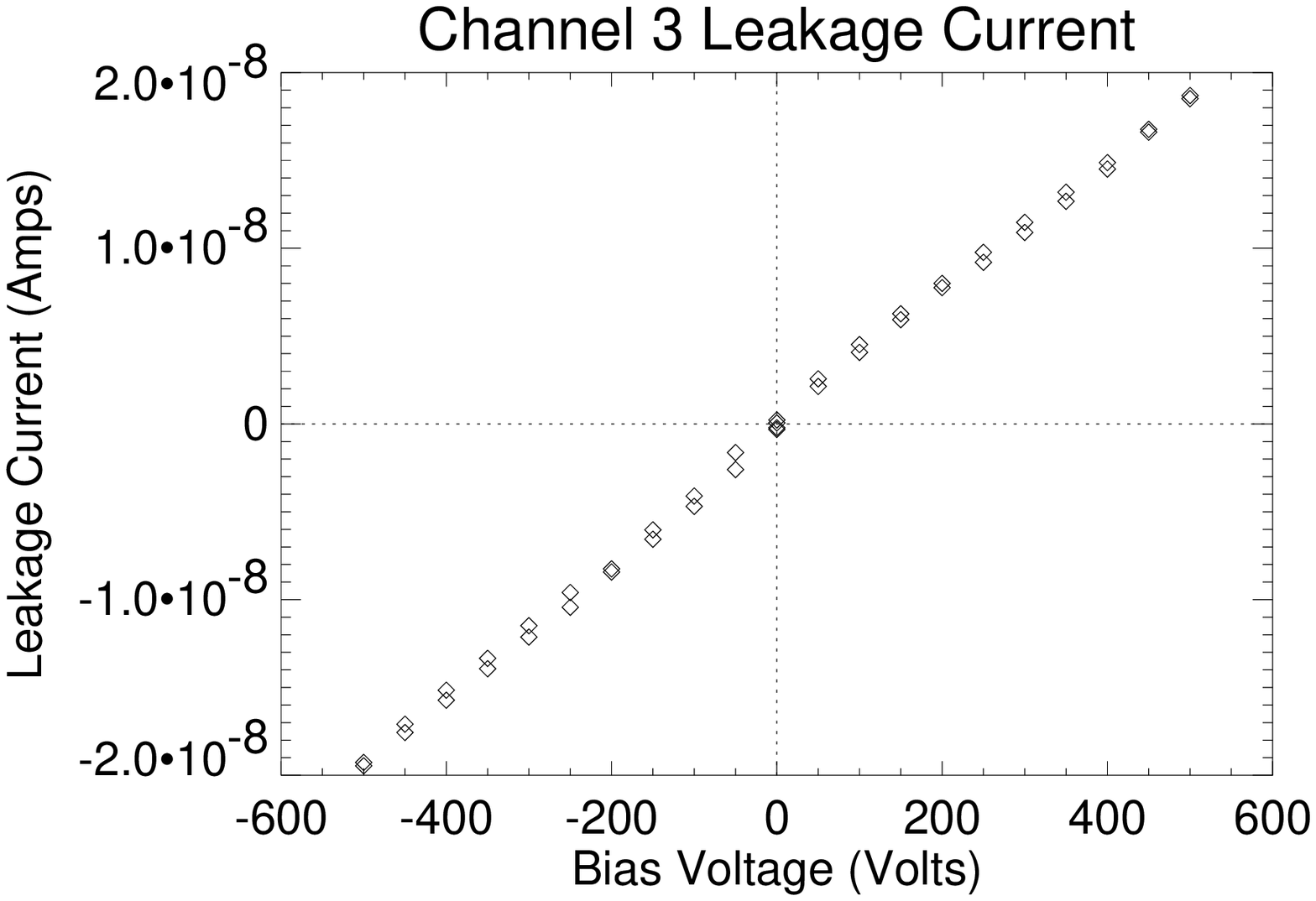,height=2.1in,width=3in}
\caption{Typical I-V curve for a pixel of the RMD array.  Typical array
operation voltage is -500 V.}
\label{fig3}
\end{minipage}
\end{figure}

The VA-1 ASIC linearity is shown in Figure~\ref{fig2}.  Voltage pulses
from the lab pulser were put into a 1.4 pF capacitor, in parallel with
the CZT input, and the peak
channel recorded in the lab MCA.  A shaping time of 1 $\mu$s was
used.  This range of input voltages
corresponds roughly to 30-180 keV in our CZT detector, and the ASIC is
found to be linear over this range.  We were limited
to this range by our MCA, but the ASIC should be linear up to 600 keV.

We measured the leakage current as a function of bias voltage for
each pixel using a Keithley 237 high voltage source and current meter.
The bias voltage was stepped from -500 V to 500 V and back down in 50 V
steps.
The detector was uniform in this respect; a typical 
I-V curve is shown in Figure~\ref{fig3}.  At our typical operating bias
voltage of -500 V, the leakage measured current is about 19 nA.  We
calculate the resistivity of the detector to be $\sim 5 \times
10^{10}$ $\Omega$-cm.

\begin{figure}[t]
\hspace{0.75in}\epsfig{file=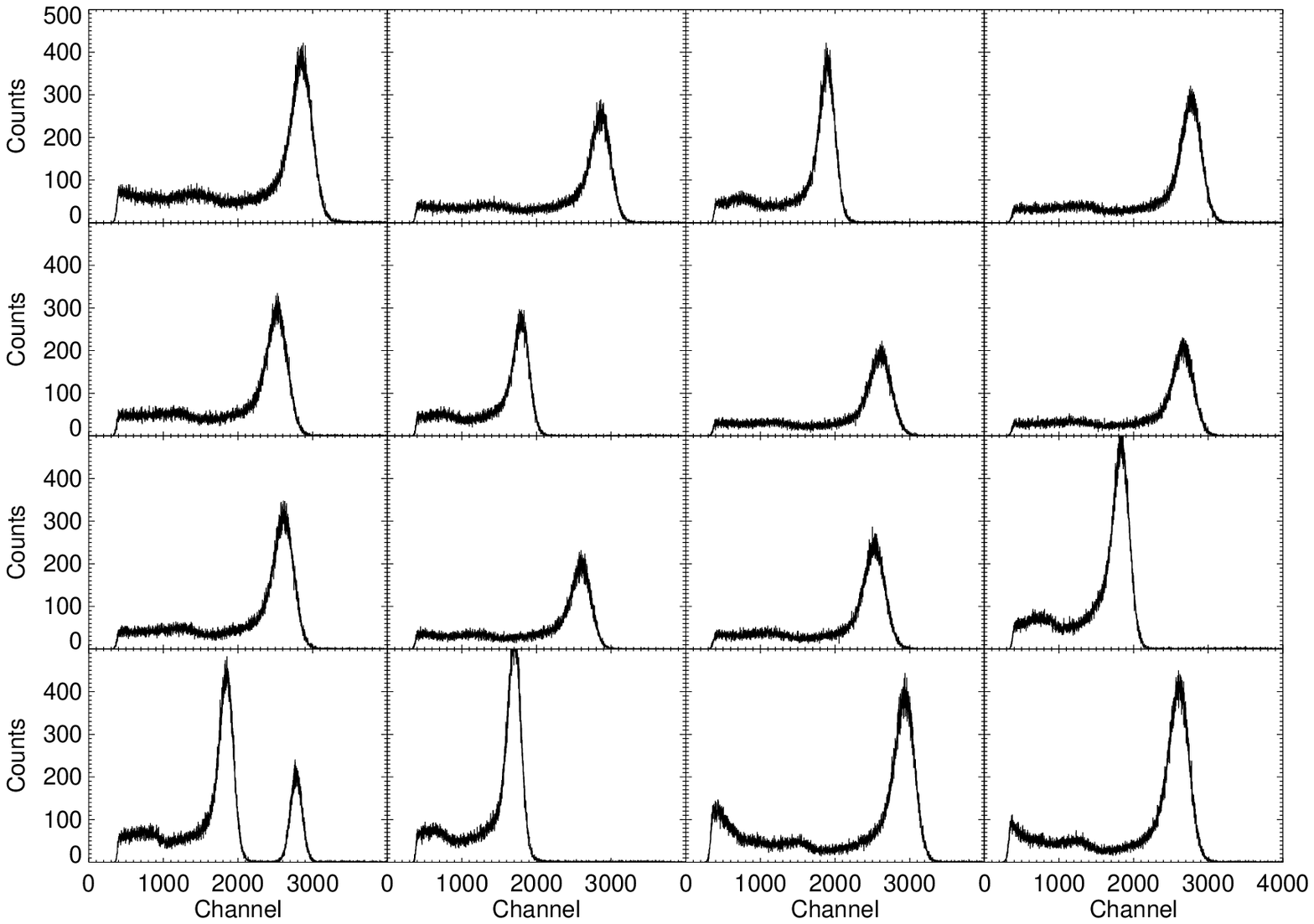,height=3.3in,width=5.0in}
\caption{$^{241}$Am spectra recorded from uniform illumination of 
all 16 channels of 
detector with -500 V bias voltage.  Pulser peak (cf. lower
left spectrum) indicates electronic noise (320e-) in the 
present simplified detector carrier and interface is dominant.}
\label{fig0}
\end{figure}

\begin{figure}[t]
\hspace{1.6in}\epsfig{file=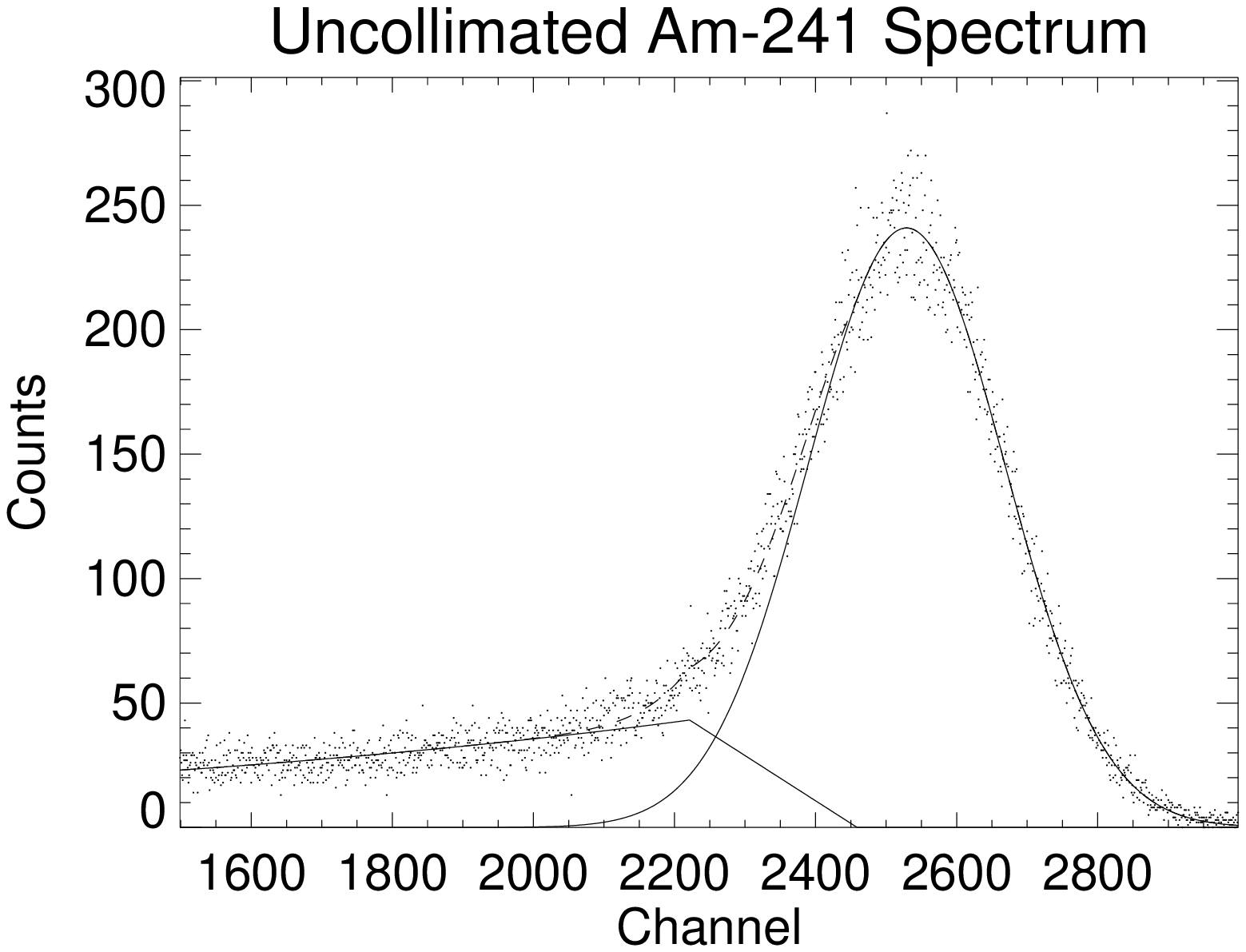,height=2.1in,width=3in}
\caption{Uncollimated $^{241}$Am spectrum from a central pixel fit by
a combination of a gaussian and low energy exponential tail. }
\label{figuncol}
\end{figure}

\begin{figure}[t] 
\begin{minipage}[t]{3.1in}
\epsfig{file=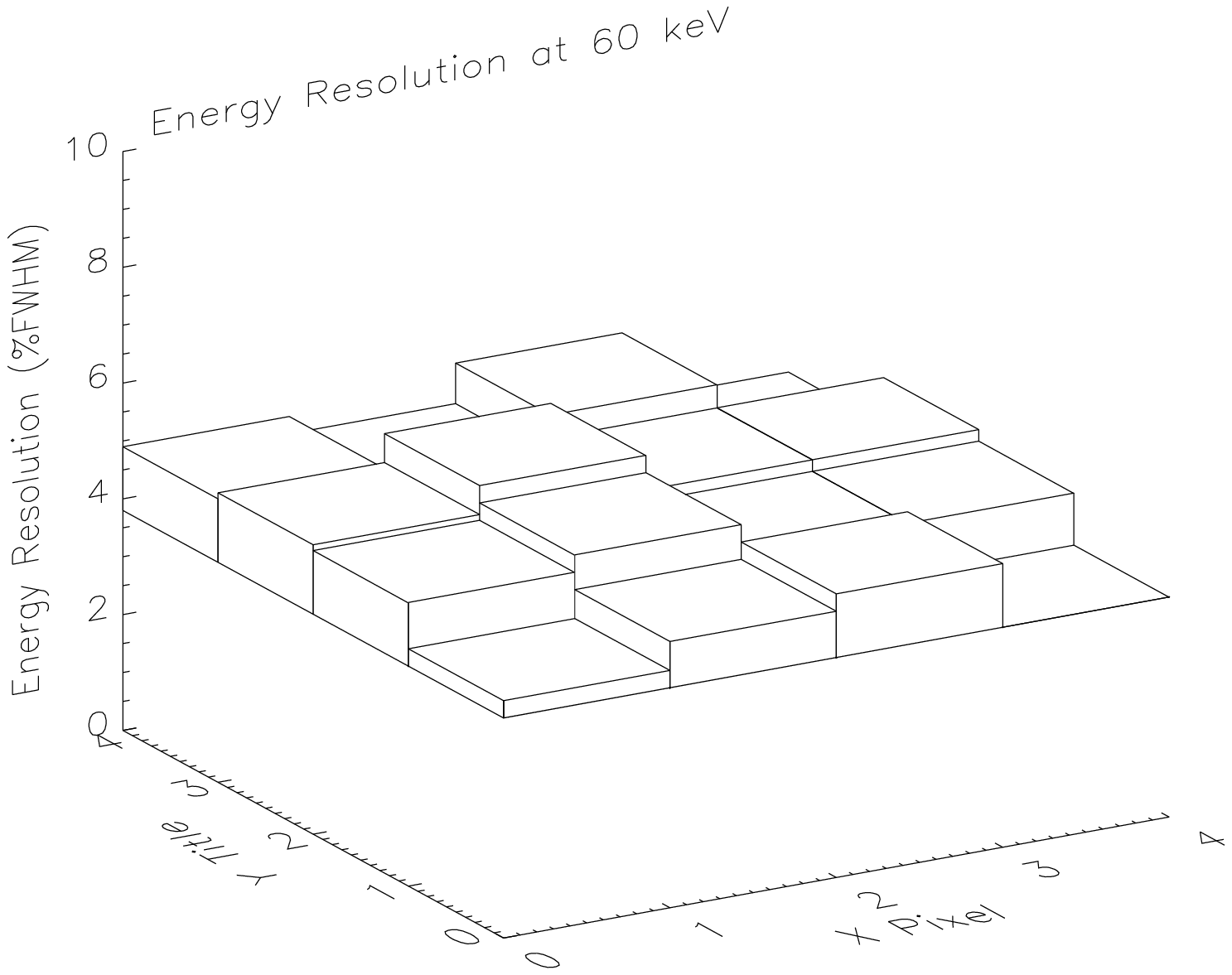,height=2.1in,width=3in}
\caption{Uniformity of detector energy resolution at 60 keV.  The FWHM
of the pulser has been subtracted in quadrature from the FWHM of the line.}
\label{fig4}
\end{minipage}
\hspace*{0.2in}
\begin{minipage}[t]{3.1in}
\epsfig{file=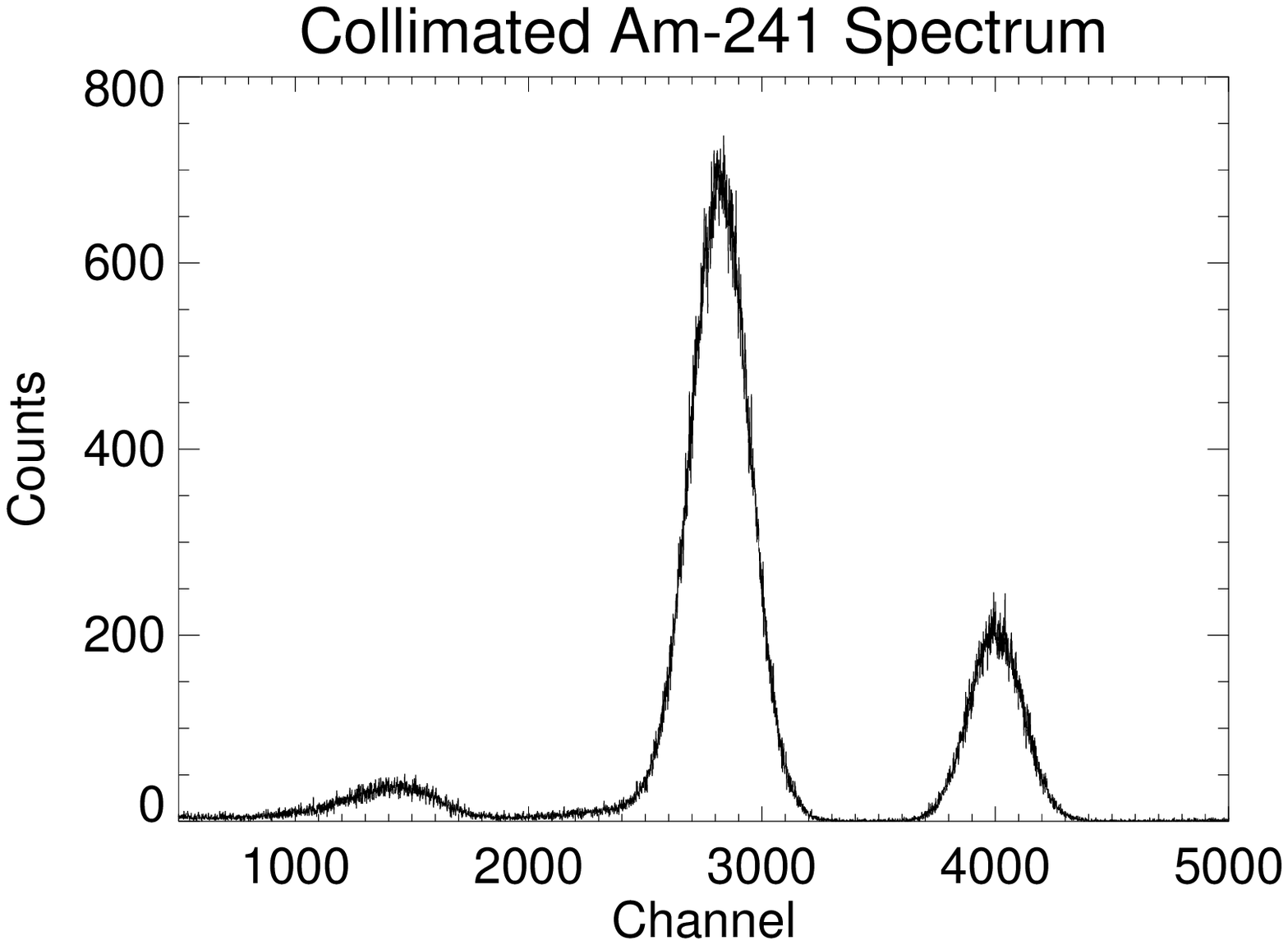,height=2.1in,width=3in}
\caption{Collimated $^{241}$Am spectrum from a central pixel with pulser
peak.  Subtracting the pulser width from the line width gives an
energy resolution of 3.8\% at 60 keV.}
\label{fig5}
\end{minipage}
\end{figure}

Spectra of 60 keV X-rays from a $^{241}$Am source were taken with the
lab MCA for all pixels.  The bias voltage was -500 V, the shaping
time was 1 $\mu$s, and the MCA was operated in a self-triggered mode
to obtain spectra of individual channels.  We also recorded spectra
with the MCA triggered by a pulse from the negatively-biased
continuous electrode, and obtained the same results.  First, spectra
were taken with the detector fully 
illuminated and a pulser injected simultaneously into all channels to
monitor electronic noise.  We obtained good spectra for all 16 pixels,
as shown in Figure~\ref{fig0}.
The low energy cutoff in these spectra is $\sim 20$ keV; the low
energy Am and Np lines are not visible in any case because the
detector is illuminated through the metal-ceramic chip carrier.
There was some variation
in the gains of the 16 channels due to differences in the line drivers on
the IDE AS test board.  The 60 keV line was fit with a
combination of a gaussian and low energy exponential tail to model the
effects of charge trapping or incomplete charge collection between
pixels, as shown in Figure~\ref{figuncol}.
We define the
photopeak efficiency as the ratio of the counts in the gaussian to the
total counts in the gaussian plus tail, and the energy resolution as the
ratio of the FWHM of the gaussian to the peak energy.  We found
typical photopeak efficiencies of 75\% for the 60 keV line, and the
energy resolution 
varied from $\sim 7-10$\%.  Most of this
peak width is due to electronic noise, as evidenced by wide pulser
peak widths.  As discussed below, the primary contribution to the
noise is probably stray capacitance in our detector carrier board, which we
know varies from channel to channel with the lengths of the wires.  We have
subtracted the pulser peak width in quadrature from the X-ray line
width to estimate the intrinsic detector energy resolution 
of $\sim 3.8 - 5.5$\%.  The pixel to
pixel variation in resolution is shown in Figure~\ref{fig4}.
Some of this variation
is still probably due to channel to channel capacitance variations.
The good photopeak efficiency and energy resolution indicate that the
detector is working in the small pixel regime.

Next we collimated the beam from the $^{241}$Am source to form a $\sim 0.25$
mm spot size on the detector and moved it from pixel to pixel with the
translation table.  The recorded spectrum of the collimated
beam on a central pixel is shown in Figure~\ref{fig5}.  The
low energy tail is practically gone; the photopeak efficiency is
97\%.  The energy resolution (with pulser width subtracted) is 3.8\%.
We took spectra of adjacent pixels to investigate the effects of
charge spreading.  The four surrounding pixels showed no counts above
background, indicating that charge spreading is localized to one pixel
even in 
a 5 mm thick detector.  Our spatial resolution is thus much finer than our
1.5 mm pixels.  We scanned the collimated beam onto the 0.2 mm interpixel
region and recorded a spectrum from the adjacent pixels.  We find here
a photopeak efficiency of 70\%, indicating that the low energy tails
originate mostly from photons incident between pixels, which naturally
results in charge division between pixels.

We have investigated the various sources of noise in our detector and
readout system by adding components one at a time and observing the
change in the pulser peak width.  The noise sources are assumed to add
in quadrature.  We find a baseline noise of the IDE AS preamp test
board and the bare readout card with the detector not in
place of $\sim 250 e^-$.  The VA-1 ASIC has a noise of $160 e^- +
10e^-/$pF of input capacitance.  This would indicate we have about $190 e^-$
noise or 19 pF of stray
capacitance in our detector carrier card and the IDE AS test board. 
Also,  we find $100e^-$ from the AC coupling
capacitors and $150e^-$ from the bias resistors. Additional noise 
contributions are $75e^-$ from the detector (chip carrier) 
capacitance and $75e^-$ from leakage current. 
We are currently making a printed circuit board detector carrier card
that should reduce the stray capacitance significantly. 

\vspace{4mm}

\noindent CONCLUSIONS

\vspace{4mm}

We have presented preliminary results from our program to develop
thick pixellated CZT array detectors with 
ASIC readouts.  For 60 keV radiation incident within a pixel, the
energy resolution and
photopeak efficiency of individual pixels are $\sim 4$\% and 97\%,
respectively.  There is
no cross-talk due to charge spreading between 
pixels, and the small pixel effect is clearly demonstrated by the near
absence of a low energy tail.
(Some effect due to charge trapping may become evident when
the electronic noise is reduced, however.)  The VA-1 ASIC performs
well, showing low noise and good linearity.  Our immediate plans
include the completion of a 16 channel parallel readout system
interfaced to a compact single board computer, which will allow true
imaging operation, and investigation of new self-triggering ASICs with
up to 64 channels to allow for larger arrays.  Balloon flight tests of
a prototype BDE array and coded aperture telescope are planned.

\vspace{4mm}

\noindent This work was supported in part by NASA grant NAG5-5103.

\vspace{4mm}

\noindent REFERENCES

\vspace{3mm}

\noindent 1. J. Butler, C. Lingren, F. Doty, Trans. Nuc. Sci. {\bf
39}, 605 (1992).

\vspace{3mm}

\noindent 2.  H. Barrett and J. Eskin, Phys. Rev. Let. {\bf 75},
156 (1995).

\vspace{3mm}

\noindent 3. A. Parsons, C. Stahle, C. Lisse, S. Babu, N. Gehrels,
B. Teegarden, P. Shu, Proc. SPIE {\bf 2305}, 121 (1994).

\vspace{3mm}

\noindent 4. C. Stahle, Z. Shi, K. Hu, S. Barthelmy, S. Snodgrass,
L. Bartlett, P. Shu, S. Lehtonen, K. Mach, Proc. SPIE {\bf 3115}, 90 (1997).

\vspace{3mm}

\noindent 5. J. Matteson, W. Coburn, F. Duttweiler, W. Heindl,
G. Huszar, P. Leblanc, M. Pelling, L. Peterson, R. Rothschild,
R. Skelton, Proc. SPIE {\bf 3115}, 160 (1997).

\vspace{3mm}

\noindent 6. T. Tumer, T. O'Neill, K. Hurley, H. Ogelman, R. Paulos,
R. Puetter, I. Kipnis, W. Hamilton, R. Proctor, in {\em The
Transparent Universe} (Proc. 2nd INTEGRAL Workshop, St. Malo, France,
1996) pp. 361-365.

\vspace{3mm}

\noindent 7.  J. Grindlay, T. Prince, N. Gehrels, J. Tueller,
C. Hailey, et al., Proc. SPIE {\bf 2518}, 202 (1995).

\end{document}